# X-ray absorption spectroscopy characterization of iron-oxide nanoparticles synthesized by high temperature plasma processing

C. Balasubramanian[1]*, B. Joseph[2], P. Gupta[3], N.L. Saini[2], S. Mukherjee[1],
D. Di Gioacchino[4] and A. Marcelli[4]

[1]FCIPT Div., Institute for Plasma Research, GIDC, Sector 25, Gandhinagar 382016, India
[2]Dip. di Fisica, Università di Roma "La Sapienza", P.le. Aldo Moro 2, 00185, Rome, Italy
[3]Centre for Converging Technologies, University of Rajasthan, Jaipur, India
[4]INFN – Laboratori Nazionali di Frascati, Via E. Fermi, 40, 00044 Frascati (RM), Italy
*Corresponding author email: balac@ipr.res.in

**Abstract**

Iron-oxide nanoparticles have been synthesized by high temperature arc plasma route with different plasma currents and characterized for their structure, morphology and local atomic order. Fe K-edge x-ray absorption spectra reveal distinct local structure of the samples grown with different plasma currents. We have shown that the local disorder is higher for the higher plasma current grown samples that also have a larger average particle-size. The results provide useful information to control structural and morphological properties of nanoparticles grown by high temperature plasma synthesis process.

**Introduction**

Materials with dimensions scaled down to the nanometer scale show novel physical and chemical properties that, however, are strongly dependent on preparation methods. This behavior is quite general for nano-structures and nanoparticles of metals, compounds and composites [1-3]. Recently, nanoparticles and nanosystems of ferric oxide have attracted large attention because of their magnetic and catalytic applications. The ferric oxide has two major crystalline phases: $\alpha$-$Fe_3O_4$ (magnetite) and $\gamma$-$Fe_2O_3$ (maghemite). These spinel-type phases have Fe in different oxidation states, i.e., both in tetrahedral and octahedral geometries. [4]. Indeed, $\alpha$-$Fe_3O_4$ is an inverse spinel ferrite in which oxygen ions form a close-packed cubic lattice with iron atoms having two different sites, i.e., tetrahedral (A) and octahedral (B) configurations. In general, the ferric oxide can be described as: $Fe^{3+}[Fe^{2+}_{1-y} Fe^{3+}_{1-y} Fe^{3+}_{1.67y} \oplus_{0.33y}]O_4$, where $y = 0$ for the $\alpha$-$Fe_3O_4$ and $y = 1$ for the $\gamma$-$Fe_3O_4$ (fully oxidized magnetite), where $\oplus$ indicates the vacancies. In the temperature range from room temperature to the Curie temperature ($T_c$ = 860 K) the A sites are populated by $Fe^{3+}$ ions and the B sites are equally populated by $Fe^{3+}$ and $Fe^{2+}$ ions as a consequence twice as many sites are populated with trivalent than with divalent iron ions [5]. It has been suggested that both bulk as well as nanoparticles of α-$Fe_3O_4$ have incomplete oxidised levels [6,7]. Indeed, the oxidation affects both structure and properties of a material and, in particular, the magnetic properties of transition metal oxides. Since oxidation depends strongly on the diffusion of oxygen atoms, the level of oxidation is highly affected by the size of nanoparticles. To a great extent, the nanoparticle size also governs the local order. Therefore, the oxidation level, and hence the properties could be tuned by different synthesis methods. Various chemical and physical processes have been used to synthesize nanoparticles of α-$Fe_3O_4$ and γ-$Fe_3O_4$ [8-13]. However, a real challenging issue of how to produce large amount of nanoparticles with a well-defined size and morphological distribution remains. Here, we report the use of a high temperature plasma route for the bulk synthesis of α-$Fe_3O_4$ nanoparticles.

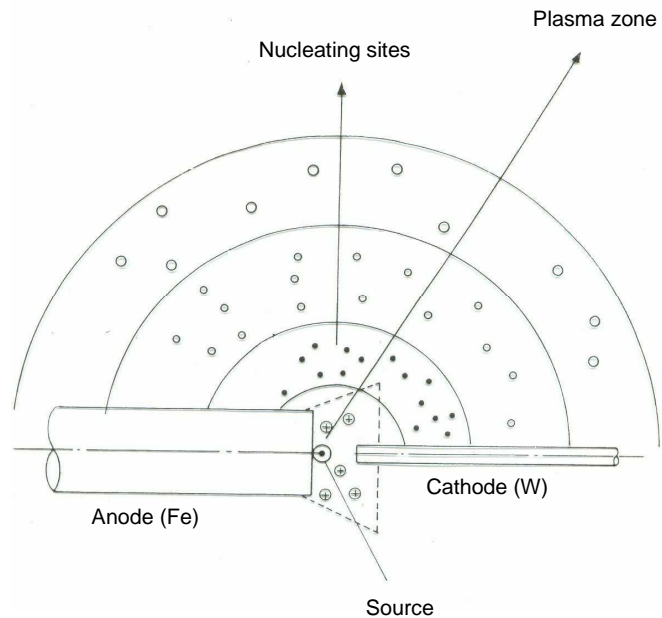

Figure 1: *Layout of the arc plasma synthesis process of nanomaterials. The grown and agglomerated nanoparticles are deposited on the substrate holder or walls of the chamber in which electrodes are placed.*

The layout of the plasma route for the synthesis of nanomaterials is outlined in Figure 1. The process involves the evaporation by high temperature, the nucleation and, finally, the condensation of the material. In the high temperature plasma synthesis, the plasma core temperature could go to thousands of Kelvin depending on the arc current. For an air plasma with currents in the range of 60 -70 A, the temperatures are expected to be in the range of 6000 - 10,000 K (i.e., electron temperature $T_e$ ~ 0.6 – 0.8 eV). The high current results, apart from increasing the evaporation rate and the particle flux, also a higher enthalpy [14] and an increased rate of thermochemical reactions [15]. However, the latter increases the probability of a multicomponent product formation [15]. During the plasma synthesis route when the products move away from the plasma zone, in addition to the high temperature, they also experience a steep temperature gradient of ~$10^4$ K/cm. This large temperature variation induces a large thermal stress that greatly affects local ordering, stoichiometry and morphology of the synthesized nanomaterials. Moreover, the large temperature variation does not provide sufficient time for the particles to grow in a highly symmetrical shapes (e.g., spheres or cubes). We have used different plasma current to synthesize iron-based nanoparticles and characterized them for their morphological properties. In addition, x-ray absorption spectroscopy has been used to study the local structure and the local order of nanoparticles of different size.

**Experimental**

A DC arc plasma has been used to generate oxide nanoparticles of iron. The electric arc was struck between the anode made with an iron metal block of diameter 40 mm (Goodfellow chemicals, purity 99.8%) and the cathode made with an iron rod of 10 mm diameter (Goodfellow Chemicals purity of 99.99+%). A multi-port, double walled stainless steel chamber water-cooled housed the electrode assembly. The cathode was a movable electrode. A constant current DC power supply was connected to the electrodes and an arc struck between them by bringing the electrodes in contact and immediately withdrawing it. The arc generated provides a high enthalpy, which in turn vaporises the electrode material. The iron metal vapours react with the oxygen in the surrounding air to form oxide. As the reacted molecules move away from the plasma zone, nucleation and cluster growth occur and the synthesized nanoparticles adhere to the inner walls of the chamber. After particles fully settle on the chamber walls, it was collected for the analysis without any post synthesis treatments. Experiments were carried out for different arc currents. We compare here materials obtained with 32 A and 65 A. The crystallinity analysis was carried out using an x-ray diffractometer (Philips, PW-1710, Cu-K radiation). The morphology was investigated using a 200 keV JEOL transmission electron microscope (JEM-1200 EX). The local atomic order and the electronic properties were determined using x-ray absorption spectroscopy (XAS) at the Fe K-edge. XAS measurements were performed at the XAFS beamline of the Elettra synchrotron radiation facility at Trieste, where the synchrotron radiation emitted by a bending magnet source operating at the energy of 2.4 GeV and with a maximum current of 140 mA, was monochromatized using a double crystal Si(111) monochromator. About 8 mg of powder samples of nanoparticles were mixed uniformly in a boron nitride matrix and pressed into pellets ~8 mm of diameter, for obtaining a unit Fe K-edge step jump at the edge. Experiments were performed at room temperature in the transmission mode using three ionization chambers mounted in series for simultaneous measurements on the sample and a thin Fe foil as the reference. Several x-ray absorption scans were measured to ensure the reproducibility of the spectra and a high signal to noise ratio. XAS data were processed using standard procedures [16] to obtain normalized x-ray absorption near edge structure (XANES) spectra and to extract the extended x-ray absorption fine structure (EXAFS) oscillation.

**Results and Discussion**

(a) TEM analysis

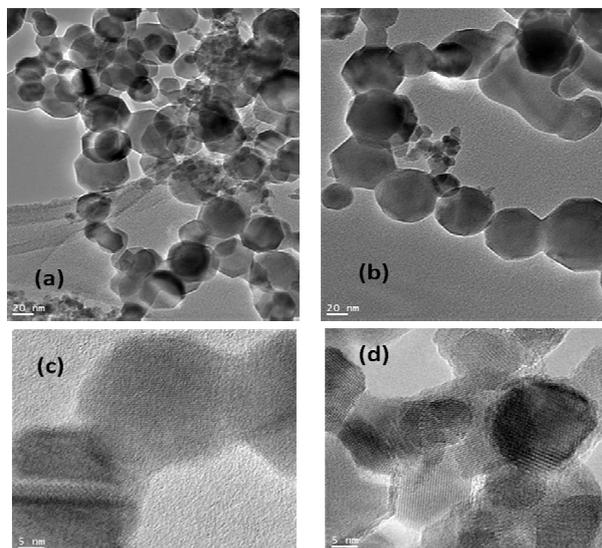

**Figure 2**: *TEM micrographs of the nanoparticles synthesised with arc currents of 32 A (panels a, c) and 65 A (panels b, d).*

Nanoparticles of both samples were dispersed on a 400 mesh copper grid and studied under a 200 keV transmission electron microscope (TEM). Figures 2 (a) and (b) display the transmission electron micrographs of the 32 A and 65 A samples, respectively. The micrographs clearly show the morphology and permits to obtain the particle size distributions. Although, shapes are slightly different, the observed variation in the particle size distribution for both samples is characteristic of high temperature plasma synthesis processes [8]. The size distribution has been calculated from a large set of TEM micrographs. Samples prepared with the lower value of current have an average particle size of ~ 34 ± 12 nm being a mixture of particles with spherical or almost octagonal shapes. For samples prepared with the arc current of 65 A, the average size is 48 ± 21 nm and shapes are mainly distorted octagons.

(b) XRD analysis

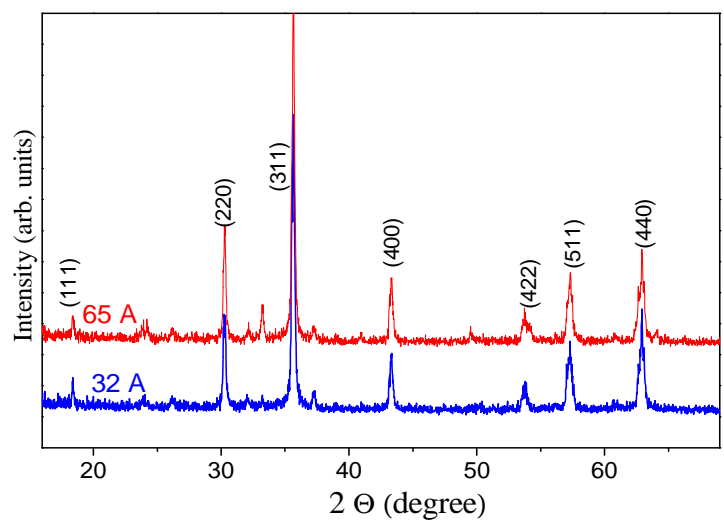

**Figure 3**: *XRD patterns of iron oxide nanoparticles synthesised under the arc currents of 32 A and 65 A.*

The crystallinity of the samples was analysed by powder x-ray diffraction (XRD). The diffraction profiles of 32 A and 65 A samples are compared in Fig. 3. There are no major differences between the two patterns except for the peak at 33.2° corresponding to a contribution due to $Fe_2O_3$ (hematite), which is better resolved in the 65 A sample. The patterns have been compared with standard XRD for the $Fe_2O_3$ (hematite) (JCPDS data: PDF number 39-1346). On the crystalline fraction, the XRD indicates possible multicomponent product formation at higher energies. Thermodynamically, depending on the enthalpy available, various thermochemical reactions are possible. Indeed, at higher enthalpy, the probability of tuning multiple reactions and products formation increases. A multiphase product formation has been already reported working at high arc current at high enthalpy [15]. However, the XRD analysis clearly indicates the presence of high crystalline phases while the "*d*" values point out the presence of both

$Fe_3O_4$ and $\gamma$-$Fe_2O_3$. The $d$ values as calculated from standard XRD pattern for $Fe_3O_4$ refers to the JCPDS data (PDF number 87-2334). The uncertainty is mainly due to the fact that the $d$ values of both phases are very close to each other and requires the use of a powerful local structural probe such as the XAS to identify and eventually discriminate among crystallographically similar phases.

(c) XAS analysis

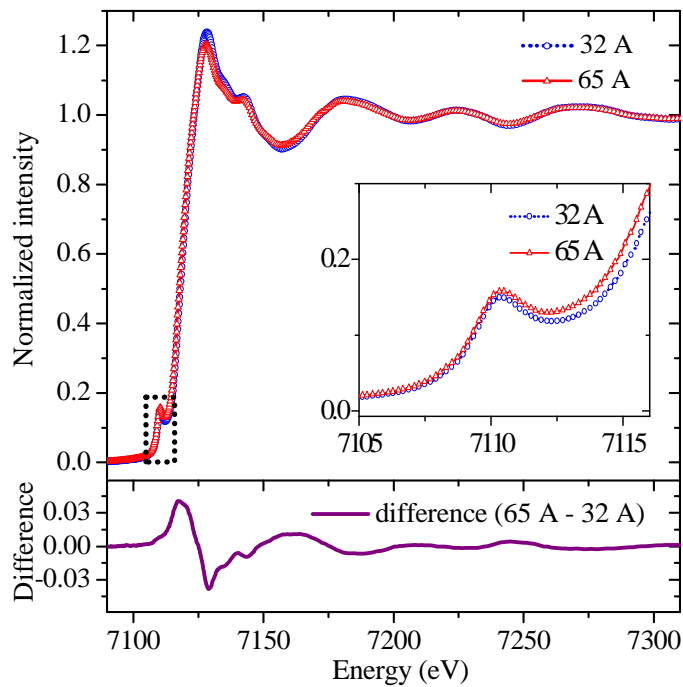

**Figure 4**: *Top panel: Fe K-edge x-ray absorption near edge structure (XANES) spectra for the 32 A (circles) and 65 A (triangles) samples. The inset shows a magnified view of the pre-edge region (dotted square in the top panel). In the lower panel the difference between the two XANES spectra is plotted showing clear differences in both pre-edge and near-edge regions.*

XAS is a site-specific local probe [16] well suited to characterize non crystalline systems and, of course, disordered systems like nanoparticles [17-20]. While, x-ray absorption near edge structure (XANES) spectra provide information on the higher order atomic correlations, i.e., bond distances and bond angles, the extended x-ray absorption fine structure (EXAFS) probes the first order atomic correlations, i.e., bond distances and coordination numbers. XANES spectra of both iron oxide samples are shown in Figure 4. A comparison between the two spectra reveals that spectral features are close to the $Fe_3O_4$ spectrum [21,22]. Therefore, both XANES and XRD data indicates that the thermal plasma-arc synthesized iron oxide nanoparticles have mainly a crystalline non stoichiometric $Fe_3O_4$ structure. Moreover, a closer inspection of the

XANES spectra points out clear differences in the spectral weights of the near-edge features. This can be better appreciated from the difference spectrum shown in the lower panel of Fig. 4. In particular, the pre-edge intensity is higher for the 65 A case (see inset in Fig. 4).

X-ray absorption is a local process in which a core electron is excited in an unoccupied electronic state via the dipole selection rule ($\Delta l = \pm 1$). The pre-edge structure in the Fe K-edge XANES (see inset in Fig. 4) are due to $1s$ to $\varepsilon d$ quadrupole allowed transitions in a distorted centro-symmetric six fold-coordinated system [23,24]. Therefore, increasing the oxidation state the position of the pre-edge peak shifts to higher energy, reflecting the crystal field splitting of $3d$ orbital sub-bands. On the other hand, intensities of the Fe K-edge pre-peak features are sensitive to the local oxygen coordination geometry. Thus, in principle, from a detailed analysis of the pre-edge features, information on the oxidation state and coordination numbers are available. In addition, the Fe K-edge XANES pre-edge peak intensity is also a measure of the local disorder [25]. In the present case, the Fe K-edge pre-edge intensity (see inset in Fig. 4) points out a reduced local order in the 65 A sample.

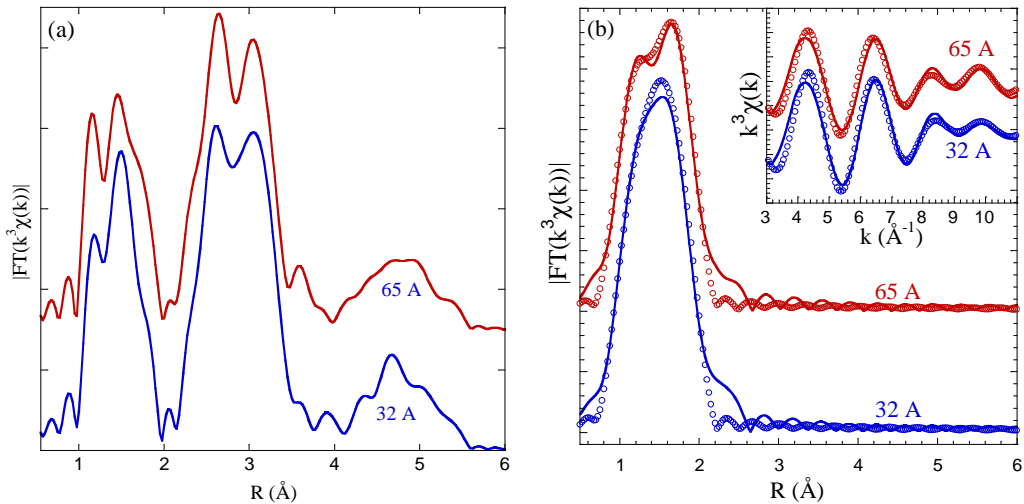

**Figure 5**: *(a) Comparison of $k^3$ weighted Fourier transform of the Fe K-edge EXAFS; (b) comparison of filtered Fourier transform and corresponding filtered EXAFS (symbols) together with models (solid lines) considering Fe-O near-neighbours: two Fe-O single scattering contributions and a multiple scattering involving the Fe-O-O path.*

A quantitative atomic distribution and information on the order can be obtained from the EXAFS analysis. Fe K-edge EXAFS has indeed used to characterize the local structure of many bulk $Fe_3O_4$ systems [21] and, in particular, on the local structural changes across the Verwey transition [26,27]. The amplitude of the Fourier transform corresponding to the EXAFS oscillations of the two nanoparticle systems is shown in Figure 5 (a). In the panel (b), we also present simulations involving Fe-O bonds corresponding to the real-space atomic distribution in the range 1-2 Å and the filtered Fourier transform amplitudes.

The local atomic distribution reported in Figure 5(a) is similar to the one in the $Fe_3O_4$ system [21,22], further confirming that nanoparticles, synthesized by the plasma-processes, have the magnetite structure. However, the intensity of the Fourier transform is higher for the 32 A sample, pointing out a different local atomic order. In general, an inverse correlation between atomic order and nanoparticles size occurs [17,18]. In our case, although the average particle size of the 65 A sample is larger, the atomic order appears lower. This can be explained as the consequence of the larger temperature gradients occurring in the synthesis of the 65 A sample. The associated large thermal stress induces also a larger local disorder of the synthesized nanoparticles.

To obtain the local structural parameters, we have analyzed the EXAFS spectra using the winXAS package [28] with the calculated phase shifts and potentials of the FEFF code [29]. For the present analysis, we have considered only two sets of oxygen near neighbours at ~1.75 Å and ~2.22 Å and modeled the filtered EXAFS oscillations. The calculated spectra describe data considering, in addition to the two Fe-O shells, also a Fe-O-O multiple scattering contribution. The modelling provides the following local atomic parameters corresponding to the oxygen atom distribution around iron atoms (for the fixed coordination number known from XRD data): 1.75 Å (0.0032 Å$^2$) and 2.22 Å (0.0032 Å$^2$) for the 35 A sample and 1.75 Å (0.0044 Å$^2$) and 2.22 Å (0.0044 Å$^2$) for the 65 A one. The values outside and inside the brackets are the bond-distances and the corresponding mean-square relative displacements (MSRD), respectively. Although local bond distances are similar for the two Fe-O scattering paths, the Fe-O-O multiple scattering contribution is quite different. The different multiple scattering contributions can be recognized by the different shapes of the amplitudes in the filtered FT of the two samples [Fig. 5 (b)]. Also the higher values of the MSRD in the 65 A case confirm the occurrence of a larger local disorder in this sample.

**Conclusion**

In this contribution, we have reported the successful synthesis of magnetite ($Fe_3O_4$) nanoparticles by a high temperature plasma synthesis process. We have found that the plasma current affects the morphology, composition and local order of iron oxide nanoparticles. The plasma synthesis permits to produce highly crystalline nanoparticles in which the degree of disorder can be tuned. The transmission electron microscopy analysis points out that, increasing the plasma current, the morphology of the nanoparticles changes from a symmetric spherical shape to a lower symmetry hexagonal shape. We have characterized the nanoparticles by Fe K-edge x-ray absorption spectroscopy that has shown clear non-stoichiometry of these $Fe_3O_4$–like nanoparticles. Furthermore, the x-ray absorption spectra reveal larger local disorder in samples synthesized with the larger plasma current that produces also nanoparticle of greater average size. The effect is a consequence of the larger temperature gradient occurring at higher plasma currents. The results provide useful feedback on bulk synthesis of metal-oxide nanoparticles with desired characteristics by plasma synthesis process.

**Acknowledgements**

The authors acknowledge the financial support of the Italy-India bilateral program on magnetic nanoparticles from the Department of Science and Technology, India and the *Bilateral Cooperation Agreement between Italy and India of the Italian Ministry of Foreign Affairs* (MAE) in the framework of the project of major relevance "*Investigating local structure and magnetism of nano-structures*". Experimental help of A. Iadecola and M. Bendele are also acknowledged.